\documentclass[a4paper,11pt]{article}
\pdfoutput=1
\usepackage[utf8]{inputenc}
\usepackage[english]{babel}
\usepackage[T1]{fontenc}
\usepackage{amsmath,amssymb}
\usepackage{array,booktabs}
\usepackage{graphicx}
\usepackage{caption}
\usepackage{subcaption}
\usepackage{amsthm}
\usepackage{bbm}
\usepackage{wasysym}
\usepackage{accents}

\theoremstyle{definition}

\usepackage{lmodern}
\usepackage[margin=2.9cm]{geometry}
\usepackage{verbatim}

\newcommand{\id}{\mathop{\mathrm{id}}}

\def\P{\ensuremath\mathcal{P}}

\usepackage{multirow}
\numberwithin{equation}{section}
\usepackage[font=small,labelfont=bf]{caption}
\usepackage{mathtools}
\usepackage{color}
\usepackage{slashed}
\usepackage{bbm}
\usepackage{braket}
\usepackage{hyperref}
\usepackage{dsfont}
\usepackage{float}
\usepackage{authblk}
\DeclareMathOperator{\tr}{tr}
\DeclareMathOperator{\Tr}{Tr}
\DeclareMathOperator{\lk}{lk}

\title{Chern-Simons Theory and the $R$-Matrix}
\author{Nanna Havn Aamand}
\affil{\textit{Perimeter Institute for Theoretical Physics, \protect\\ Waterloo, ON N2L 2Y5, Canada}}
\date{}

\begin{document}
\maketitle
\begin{abstract} 
It has been a long-standing problem how to relate Chern-Simons theory to the quantum groups. In this paper we recover the classical $r$-matrix directly from a 3-dimensional Chern-Simons theory with boundary conditions, thus creating a direct link to the quantum groups. It is known that the Jones polynomials can be constructed using an $R$-matrix. We show how these constructions can be seen to arise directly from 3-dimensional Chern-Simons theory.
\end{abstract}

\section{Introduction}
It was first shown by Witten in a famous paper \cite{witten1989quantum} that the expectation value of Wilson loops in 3-dimensional Chern-Simons theory gives rise to certain values of the Jones polynomials of knots. On the other hand Reshetikhin and Turaev \cite{turaev1988yang,reshetikhin1988quantized} have given constructions of the Jones polynomials from quantum groups, by using an $R$-matrix representation of the Artin braid group. Until now, it has however been unclear how the constructions of Reshetikhin and Turaev can be seen to arise directly from 3-dimensional Chern-Simons theory. The aim of the present paper is to fill in this gap. Motivated by recent papers by Costello, Witten and Yamazaki~\cite{costello2017gauge,costello2018gauge} we show, working to leading order in perturbation theory, that the propagator of a 3-dimensional Chern-Simons theory with gauge group $G=SL_2(\mathbb{C})$ has the form of an $R$-matrix when imposing boundary conditions that break the $G$-symmetry of the action.  This result allows us to give an explicit construction of the Jones polynomials from the expectation value of Wilson loops in the theory. In fact, by choosing a gauge where interactions through the $R$-matrix only occur at the points where two Wilson lines cross in $\mathbb{R}^2$, we obtain a Hecke algebra representation of the Artin braid group with Wilson lines interpreted as braid strands. The original construction of the Jones two-variable polynomials \cite{jones1990hecke} comes from a Markov trace due to Ocneanu \cite{freyd1985new} acting on a Hecke algebra representation of the Artin braid group. We show that the expectation value of Wilson loops obtained from the closure of Wilson lines behaves like Ocneanu's trace function and thus it can be normalized to give the Jones polynomials for specific values of the variables. \\

Guadagnini et al. \cite{guadagnini1990wilson,cotta1990quantum,guadagnini1990chern,guadagnini1991link} have similarly studied the problem of recovering link polynomials from perturbative Chern-Simons theory. It was argued in \cite{guadagnini1991link} that, without breaking the $G$-symmetry, one recovers the $R$-matrix of a quasi-triangular quasi-Hopf algebra, and in \cite{guadagnini1990chern} that Wilson line operators are related to a monodromy representation of the braid group. However, until now no explicit construction of the $R$-matrix has been made. The approach of Guadagnini et al. was further studied by Morozov and Smirnov \cite{morozov2010chern} using a temporal gauge condition. However, since the $G$-symmetry of the Chern-Simons action is not broken they do not recover the $R$-matrix. 

\section{The $R$-Matrix}\label{sec:YBE}
In this section we briefly review the Yang-Baxter formalism \cite{yang1967some,baxter1971eight} and present the solutions of the classical Yang-Baxter equation, $r\in \mathfrak{g}\otimes\mathfrak{g}$, for the Lie algebra $\mathfrak{g}=\mathfrak{sl}_2(\mathbb{C})$ which we will be considering in the rest of the paper.\\

\begin{figure}[H]
 	\centering
    \includegraphics[width=0.32\paperwidth]{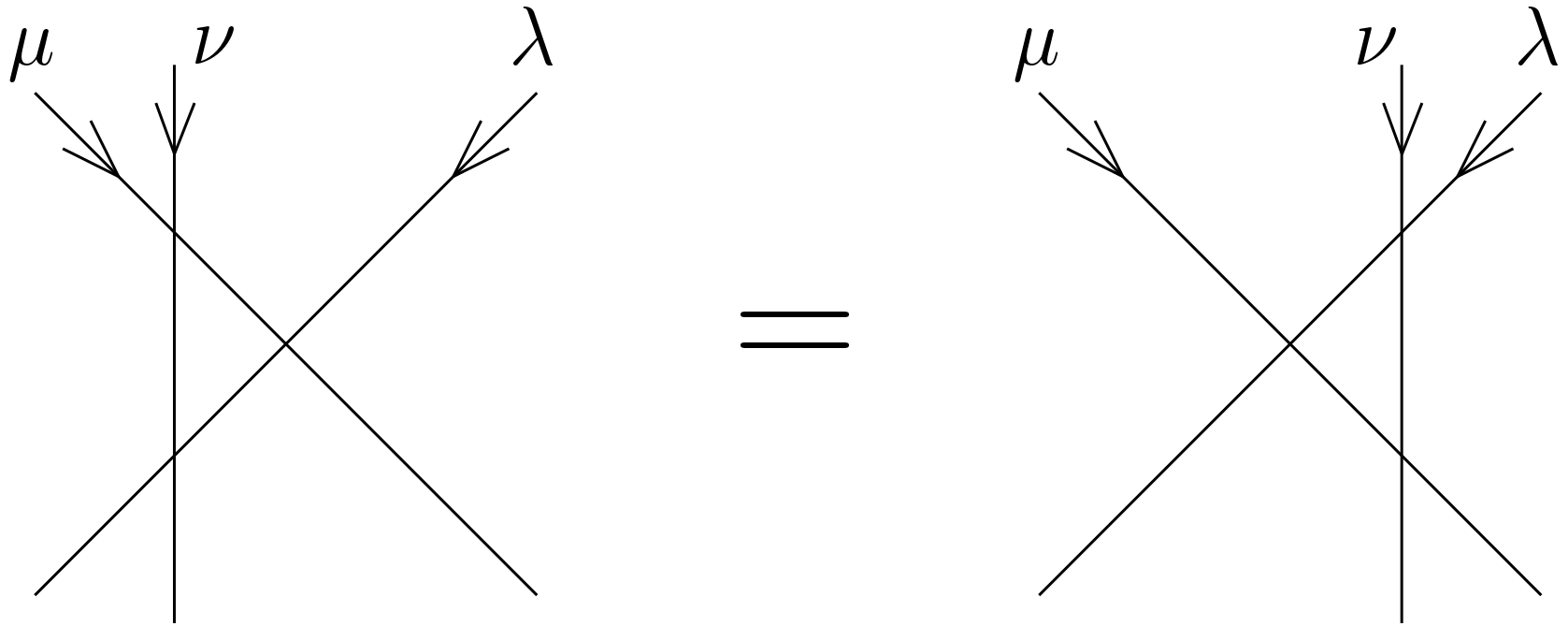}
    \caption{Graphical representation of the Yang-Baxter equation.}
    \label{fig:YBE}
\end{figure}

For an $n$-dimensional vector space $V$, let $R$ be a bilinear operator, $R:V\otimes V\to V\otimes V$. Furthermore, consider $k$ copies of $V$ labeled by $V_1,\dots,V_k$ and define $R_{\mu\nu}:V^{\otimes k}\to V^{\otimes k}$, $\mu,\nu\in\{1,\dots,k\}$ to be the operator obtained by first acting with $R$ on $V_\mu$ and $V_\nu$ and then acting with the permutation operator $P_{\mu\nu}:V^{\otimes k}\to V^{\otimes k}$ that swaps a vector from $V_\mu$ and a vector from $V_\nu$. For example for $k=3$ we have,
\begin{align}
R_{12}=P_{12}(R\otimes \id) :V_1\otimes V_2\otimes V_3\to V_2\otimes V_1\otimes V_3.
\end{align}
$R$ is said to be an $R$-matrix if it satisfies the relation,
    \begin{align}
        R_{\mu\nu}R_{\mu\lambda}R_{\nu\lambda}=R_{\nu\lambda}R_{\mu\lambda}R_{\mu\nu} \label{eq:YBE}
    \end{align}  
known as the Yang-Baxter equation. The Yang-Baxter equation is most easily understood from a graphical representation, as the one given in Figure~\ref{fig:YBE}. 
\noindent In Figure~\ref{fig:YBE} each line represents a vector space, $V_\mu$, $V_\nu$ and $V_\lambda$, and each crossing between two lines represents an $R$-matrix followed by a permutation  acting on the corresponding vector spaces. In this picture, the Yang-Baxter equation \eqref{eq:YBE} tells us that the middle line can be pulled across the crossing between two other lines without changing the total outcome.

\subsection{The Classical $r$-Matrix} \label{sec:CYBE}
In the following we study solutions of the Yang-Baxter equation in the context of gauge theories. We therefore consider a gauge group $G$ with Lie algebra $\mathfrak{g}$, and take $R$ to be an element of $\mathfrak{g}\otimes\mathfrak{g}$. The vector space $V$ then corresponds to the space of spin states of $\mathfrak{g}$. Since we will be working to leading order in perturbation theory, we furthermore write $R$ as an expansion around the identity in the expansion parameter $\hbar$, as $R=I+\hbar r+\mathcal{O}(\hbar^2)$. Here $r\in \mathfrak{g}\otimes\mathfrak{g}$ is known as the classical $r$-matrix. Inserting this expansion into equation \eqref{eq:YBE} we find from the terms at order $\mathcal{O}(\hbar^2)$ that $r$ must satisfy the following equation, known as the classical Yang-Baxter equation,
\begin{align}
    [r_{\mu\nu},r_{\mu\lambda}]+[r_{\mu\nu},r_{\nu\lambda}]+[r_{\mu\lambda},r_{\nu\lambda}]=0 .\label{CYBE}
\end{align}
As previously mentioned, we will in the present paper work with the gauge group $G=SL_2(\mathbb{C})$, with corresponding Lie algebra $\mathfrak{g}=\mathfrak{sl_2(\mathbb{C})}$ consisting of all traceless 2 by 2 matrices. The basis elements $e,f,h$ of $\mathfrak{sl}_2(\mathbb{C})$ in the fundamental representation are given by
\begin{align}
    e=\begin{pmatrix}
    0 & 1 \\ 0 & 0  
    \end{pmatrix}\hspace{15pt} 
    f=\begin{pmatrix}
    0 & 0 \\ 1 & 0 \end{pmatrix} \hspace{15pt} h=\begin{pmatrix}
    1 & 0 \\ 0 & -1 \end{pmatrix}, \label{eq:generators}
\end{align}
from which we can infer the Lie brackets:
\begin{align}
    [e,f]=h, \hspace{15pt} [h,e]
    =2 e ,\hspace{15pt}[h,f]=-2f . \label{Lie brack}
\end{align}
The solutions of the classical Yang-Baxter equation for $\mathfrak{g}=\mathfrak{sl}_2(\mathbb{C})$ can be found in \cite{chari1995guide}. We~have
\begin{align}
    r=e\otimes f +\frac{1}{4}h\otimes h \ . \label{CPsol1}
\end{align}
Another solution can be obtained by interchanging $e$ and $f$ since this only changes the left side of \eqref{CYBE} by an overall minus sign. We get
\begin{align}
    \tilde{r}=f\otimes e +\frac{1}{4}h\otimes h \ . \label{CPsol2}
\end{align}
Notice that the solutions \eqref{CPsol1} and \eqref{CPsol2} do not have full $SL_2(\mathbb{C})$-symmetry. Since we hope to recover these solutions from Chern-Simons theory we must therefore find a way of breaking the $SL_2(\mathbb{C})$-symmetry of the Chern-Simons action. As we shall see in the next section, this can be done by imposing specific boundary conditions to the gauge field. 
\section{Chern-Simons Theory and the $R$-Matrix} \label{sec:CS-theory}
In this section we show how the classical Yang-Baxter solutions presented in Section~\ref{sec:CYBE} can be recovered from 3-dimensional Chern-Simons theory. More concretely, we will consider the usual 3-dimensional Chern-Simons theory defined on the manifold $M=\mathbb{R}\times I$, where $I$ is a closed interval. The Chern-Simons action has the form
\begin{align}
    S_{\text{CS}}&=\frac{1}{4\pi}\int_{\mathbb{R}^2\times I} \Tr\left(A\wedge dA+\frac{2}{3}A\wedge A\wedge A\right). \label{CS-action}
\end{align}
Let $x_1$, $x_2$ be coordinates on $\mathbb{R}^2$ and $x_3$ a coordinate on $I$, then $A~=~A_1dx_1+A_2dx_2+A_3dx_3$, where the $A_i$'s are elements of the Lie algebra $\mathfrak{g}$ of the gauge group. $\Tr$ denotes a non-degenerate invariant bilinear form on $\mathfrak{g}$. In the case of $\mathfrak{g}=\mathfrak{sl}_2(\mathbb{C})$ in the fundamental 2~by~2 representation  \eqref{eq:generators}, we can take $\Tr$ to be the usual trace: $\Tr(ef)=1$, $\Tr(hh)=2$.\\

We show in the following that by imposing boundary conditions on the gauge field at the endpoints of $I$ consistent with those proposed in \cite{costello2017gauge}, the propagator of the Chern-Simons action \eqref{CS-action} gives the classical Yang-Baxter solutions \eqref{CPsol1} and \eqref{CPsol2}.

\subsection{Boundary Conditions} \label{sec:BC}
Since the the Chern-Simons action is only gauge invariant up to a surface term, we must make sure that the this term vanishes with the chosen boundary conditions. Under a gauge transformation $A\to A+\delta A$ where $\delta A$ is an exact one-form, the variation of the Chern-Simons action is given by, 
\begin{align}\label{eq:var}
\delta S_{\text{CS}}=\frac{1}{2\pi}\int_{\mathbb{R}^2\times \partial I}\Tr A \wedge \delta A  ,
\end{align} 
It was argued in \cite{costello2017gauge} (in the case of a 4-dimension generalisation of the usual Chern-Simons action) that, in order to make the boundary term of the Chern-Simons action vanish while reproducing a solution of the Yang-Baxter equation, one must choose the boundary conditions as follows: For a given Lie algebra $\mathfrak{g}$, let $\mathfrak{l}_0$ and $\mathfrak{l}_1$ be middle-dimensional subalgebras of $\mathfrak{g}$ on which $\Tr({\cdot,\cdot})$ vanishes and which satisfy $\mathfrak{l}_0\cap\mathfrak{l}_1=0$ (or equivalently $\mathfrak{l}_0\oplus\mathfrak{l}_1=\mathfrak{g}$). Choosing for convenience $I=[0,1]$, we then require $A$ and $\delta A$ to take value in $\mathfrak{l}_0$ on the boundary $\mathbb{R}^2\times \{0\}$ and in $\mathfrak{l}_1$ on the boundary $\mathbb{R}^2\times \{1\}$. Clearly, it is not possible to construct such $\mathfrak{l}_0$ and $\mathfrak{l}_1$ for $\mathfrak{g}=\mathfrak{sl}_2(\mathbb{C})$, since this algebra has odd dimension. We will therefore (following \cite{costello2017gauge}) extend the dimension of the algebra by 1, adding to $\mathfrak{sl}_2(\mathbb{C})$ another copy $\tilde{h}$ of the Cartan $h$ of $\mathfrak{g}$. The resulting Lie algebra thus becomes $\mathfrak{g}=\mathfrak{sl}_2(\mathbb{C})\oplus \tilde{h}$. We extend the invariant bilinear form on $\mathfrak{sl}_2(\mathbb{C})$ to $\mathfrak{g}$ by defining $\Tr({\tilde{h},\tilde{h}})=2$ and $\Tr(\tilde{h},a)=0$ for all $a\in\mathfrak{sl}_2(\mathbb{C})$. The required properties for $\mathfrak{l}_0$ and $\mathfrak{l}_1$ can now be satisfied by setting $\mathfrak{l}_0=f\oplus (h-i\tilde{h})$ and $\mathfrak{l}_1=e\oplus (h-i\tilde{h})$. We therefore arrive at the following boundary conditions on the gauge field: 
\begin{equation}
\begin{aligned}\label{eq:bc}
\mathbb{R}^2\times\{0\}&:\hspace{20pt} A^{e}_{i}=0 \hspace{20pt} A^{h}_{i}+iA_i^{\tilde{h}}=0\\
\mathbb{R}^2\times\{1\}&:\hspace{20pt} A^{f}_{i}=0 \hspace{20pt} A^{h}_{i}-iA_i^{\tilde{h}}=0 \ .
\end{aligned}
\end{equation} 
We will in the following take $\tilde{h}$ to act as the identity in the fundamental representation given in Section~\ref{sec:CYBE}:
\begin{align}\label{eq:htilde}
\tilde{h}=\begin{pmatrix}
1&0\\0&1
\end{pmatrix} .
\end{align}
We now proceed to determining the propagator of the theory in the presence of these boundary conditions.
 
\subsection{The Propagator}\label{sec:Prop}
The easiest way to compute the propagator of the theory with boundary conditions is by first computing the propagator of the free theory (with no boundary conditions) and then modifying it so that the boundary conditions are satisfied. This will be done in the following.
\subsubsection{The Propagator of the Free Theory}
The propagator, interpreted as a 2-form in the variables $x$ and $x'$, has the form, 
\begin{align}
P^{ab}(x,x')=\sum_{i,j=1,2,3}\braket{A_i^a(x),A_j^b(x')}\mathrm{d}(x^i-{x'}^i)\wedge \mathrm{d}(x^j-{x'}^j),
\end{align}
where $a,b\in\{e,f,h,\tilde{h}\}$ are color indices. The expression becomes particularly simple if we choose as our gauge the following modified version of the Lorentz gauge\footnote{This gauge condition follows from the Lorentz gauge by rescaling the $x_1$ and $x_2$ components of the gauge field, $A_1'=\lambda^{-1} A_1$, $A_2'=\lambda^{-1} A_2$, $A_3'=A_3$, and then taking the limit $\lambda\to 0$. We are allowed to do this since the theory is metric independent.}, 
\begin{align}
    \partial_{x_3}A_{3}=0. \label{gauge}
\end{align}
In this gauge the propagator 2-form $P^{ab}_{ij}(x,x')\coloneqq\braket{A^a_i(x),A^b_j(x')}$ is defined through the relations,
\begin{equation}
\begin{aligned}\label{eq:props}
    &\partial_{x_3}P^{ab}_{3j}(x,x')=0 \ , \ j\in\{1,2,3\} \\
    &\Tr(t^a t^b)\mathrm{d}P^{ab}(x,x')=4\pi\hspace{1pt}\delta_{x_1=x_1'}\delta_{x_2=x_2'}\delta_{x_3=x_3'},
\end{aligned}
\end{equation}
along with the anti-symmetry property $P^{ab}(x,x')=-P^{ba}(x',x)$, which follows from the anti-symmetry of the kinetic term in the Chern-Simons action. The second equation in \eqref{eq:props} implies that the color dependence of the propagator is given by the quadratic Casimir of the Lie algebra $\mathfrak{g}$. Thus, if we reinterpret the propagator to be an element of $\mathfrak{g}\otimes\mathfrak{g}$ it takes the form $P(x,x')C(\mathfrak{g})$, where $C(\mathfrak{g})$ is the quadratic Casimir of $\mathfrak{g}=\mathfrak{sl}_2(\mathbb{C})\oplus\tilde{h}$:
\begin{align}
    C(\mathfrak{g})=e\otimes f +f\otimes e+\frac{1}{2}h\otimes h+\frac{1}{2}\tilde{h}\otimes\tilde{h}. 
\end{align}
It can easily be verified that the conditions in \eqref{eq:props} are satisfied if we choose as our ansatz the following expression for the propagator,
\begin{align}
    P=2\pi\delta_{x_1=x_1'}\delta_{x_2=x_2'}\big(\delta_{x_3>x_3'}-\delta_{x_3< x_3'}\big) C(\mathfrak{g}). \label{free.prop}
\end{align}
Thus, we have determined the propagator of the free theory and we are ready to impose the boundary conditions discussed in Section~\ref{sec:BC}. 
\subsubsection{The Propagator with Boundary Conditions}
The chosen set of boundary conditions \eqref{eq:bc} translates into the following constraints on the propagator: In the case of $x_3 < x_3'$ we have
\begin{align}
    P^{ea}(x_1,x_2,x_3=0,x')&=P^{af}(x,x_1',x_2',x_3'=1)=0 \nonumber \\ P^{ha}(x_1,x_2,x_3=0,x')&=-P^{\Tilde{h}a}(x_1,x_2,x_3=0,x') \label{const1}\\
    P^{ah}(x,x_1',x_2',x_3'=1)&=P^{a\Tilde{h}}(x, x_1',x_2',x_3'=1)
    \nonumber
\end{align}
for any $a\in\{e,f,h,\Tilde{h}\}$, and in the case of $x_3>x_3'$ we have
\begin{align}
    P^{fa}(x_1,x_2,x_3=1,x')&=P^{ae}(x,x_1',x_2',x_3'=0)=0 \nonumber \\P^{ah}(x,x_1',x_2',x_3'=0)&=-P^{a\Tilde{h}}(x,x_1',x_2',x_3'=0) \label{const2}\\ P^{ha}(x_1,x_2,x_3=1,x')&=P^{\Tilde{h}a}(x_1,x_2,x_3=1,x') \ .\nonumber
\end{align}
Since the propagator in \eqref{free.prop} obviously has translation invariance, the constraints in \eqref{const1}, previously evaluated at $x_3=0$ and $x_3'=1$, must actually hold for all $x_3$ and $x_3'$ with $x_3<x_3'$. Similarly, the constraints in \eqref{const2} must hold for all $x_3$ and $x_3'$ with $x_3>x_3'$. Thus we can write the total constraints on the propagator imposed by the boundary conditions as follows:
\begin{equation}
\begin{aligned}\label{eq:fullconst}
    &x_3<x_3': \hspace{20pt}P^{ea}=P^{af}=0 , \ \  P^{ha}=-P^{\Tilde{h}a} \ , \ \ 
    P^{ah}=P^{a\Tilde{h}}\\
    &x_3>x_3': \hspace{20pt}P^{fa}=P^{ae}=0 , \ \  P^{ah}=-P^{a\Tilde{h}} \ , \ \ 
    P^{ha}=P^{\Tilde{h}a}\ . 
\end{aligned}
\end{equation}
Starting from the free propagator $P$ in \eqref{free.prop} we can construct a propagator in the presence of boundary conditions by adding a term $P'$ that compensates for the relevant elements of $P$ such that \eqref{eq:fullconst} is satisfied. In order for the result to still be a valid propagator, $P'$ must satisfy the gauge condition \eqref{gauge}, have vanishing exterior derivative, and obey the anti-symmetry property $P'_{ab}(x,x')=-P'_{ba}(x',x)$.  Going back to the formalism of \eqref{free.prop} where the propagator is taken to be an element of $\mathfrak{g}\otimes\mathfrak{g}$, we define 
\begin{align}
    P'=2\pi\hspace{1pt}\delta_{x_1=x_1'}\delta_{x_2=x_2'}\Big(\delta_{x_3>x_3'}+\delta_{x_3<x_3'}\Big)\Big(e\otimes f-f\otimes e+\frac{i}{2}\Tilde{h}\otimes h -\frac{i}{2}h\otimes\Tilde{h}\Big) ,
\end{align}
which has the required properties. By adding $P'$ to the free propagator we reach the following expression for the propagator in the theory with boundary conditions
\begin{equation}
\begin{aligned}
    P\to P+P'=4\pi\hspace{1pt}&\Big(e\otimes f+\frac{1}{4}(h+i\tilde{h})\otimes(h-i\tilde{h})\Big)\delta_{ x_1=x_1'}\delta_{x_2=x_2'}\delta_{x_3> x_3'}\\&-4\pi\hspace{1pt}\Big(f\otimes e+\frac{1}{4}(h-i\tilde{h})\otimes(h+i\tilde{h})\Big)\delta _{x_1=x_1'}\delta_{x_2=x_2'}\delta_{x_3<x_3'}  . \label{prop}
\end{aligned}
\end{equation}
Let us compare the color factors in this result with the solutions for the classical $R$-matrix given in \eqref{CPsol1}, \eqref{CPsol2}. Since $\Tilde{h}$ commutes with all the generators of $\mathfrak{sl_2}(\mathbb{C})$, one easily finds that the Yang-Baxter equation is still satisfied if we include $\Tilde{h}$ in the solutions $r$ and $\tilde{r}$ as in \eqref{prop}. Thus, we can rewrite the propagator as,
\begin{align}
P(x,x')=4\pi\hspace{1pt} \delta_{ x_1=x_1'}\delta_{x_2=x_2'}\big(r\hspace{1pt}\delta_{x_3> x_3'}-\Tilde{r}\hspace{1pt}\delta_{x_3<x_3'}\big),\label{eq:propagator}
\end{align}
We have thus managed to recover solutions of classical Yang-Baxter equation from a 3-dimensional Chern-Simons theory by using the approach suggested in \cite{costello2017gauge,costello2018gauge}. As mentioned in the introduction, Turaev and Reshitikhin have previously given constructions of the Jones polynomials using an $R$-matrix representation of the Artin braid group. The purpose of the remaining part of this paper will be to explain how these construction arise from Chern-Simons theory. Our first step towards this is to introduce Wilson lines to the theory since they, as we will argue in the following, can be seen as representing braid strands.

\section{Wilson Loops and Knots}\label{sec:WL}
In this section we study one of the fundamental gauge invariant observables of Chern-Simons theory known as Wilson loops. A Wilson loop is obtained by taking the trace of the holonomy of the gauge field $A$ around a simple, smooth, closed curve $\gamma:[0,1]\to\mathbb{R}^2\times I$,
\begin{equation}
\begin{aligned}\label{eq:W1}
W(\gamma)&=\Tr\mathcal{P}\exp\left(\oint_\gamma A\right)\\ &=\Tr(\mathds{1})+\Tr\oint_\gamma dx^i A_i(x)+\Tr\oint_{\gamma}dx^i\int^xdx'^j A_i(x)A_j(x')+\dots
\end{aligned}
\end{equation}
where $\mathcal{P}$ stands for the path ordering of the exponential. In the context of the present paper, the trace is taken over the fundamental representation of the gauge group $\mathfrak{g}=\mathfrak{sl}_2\oplus \tilde{h}$ with generators $\{e, f, h, \tilde{h}\}$ given in \eqref{eq:generators} and \eqref{eq:htilde}. \\

A single Wilson loop, represented by a closed, oriented curve $\gamma$ in $\mathbb{R}^2\times I$, is called a \textit{knot} and a collection of Wilson loops given by $n$ closed, oriented curves $\{\gamma_1,\dots,\gamma_n\}$ that are non-intersecting but may be linked around each other is called an ($n$-component) \textit{link}. In the following sections we will be concerned with studying the expectation value of such links, which we will write as $\braket{W(L)}=\braket{W(\gamma_1)\dots W(\gamma_n)}$. 

\subsection{Interacting Wilson Lines}\label{sec:IntWL}
We start by studying the interaction of open Wilson lines, which means that we will for now be ignoring the trace in \eqref{eq:W1}.\\

For any set of Wilson lines, the leading order interaction between them is given by the propagator in \eqref{eq:propagator} and thus is only non-vanishing at the points where two lines cross in $\mathbb{R}^2$. We will therefore in the following represent Wilson lines by their planar projection onto $\mathbb{R}^2$. In this representation each line corresponds to a space of spin states $V_\mu$ of the fundamental representation of $\mathfrak{sl}_2\oplus\tilde{h}$, and to each crossing is attached an interaction-matrix, given by the propagator, which acts on the corresponding vector spaces.
\begin{figure}[H]
 	\centering
    \includegraphics[width=0.3\paperwidth]				{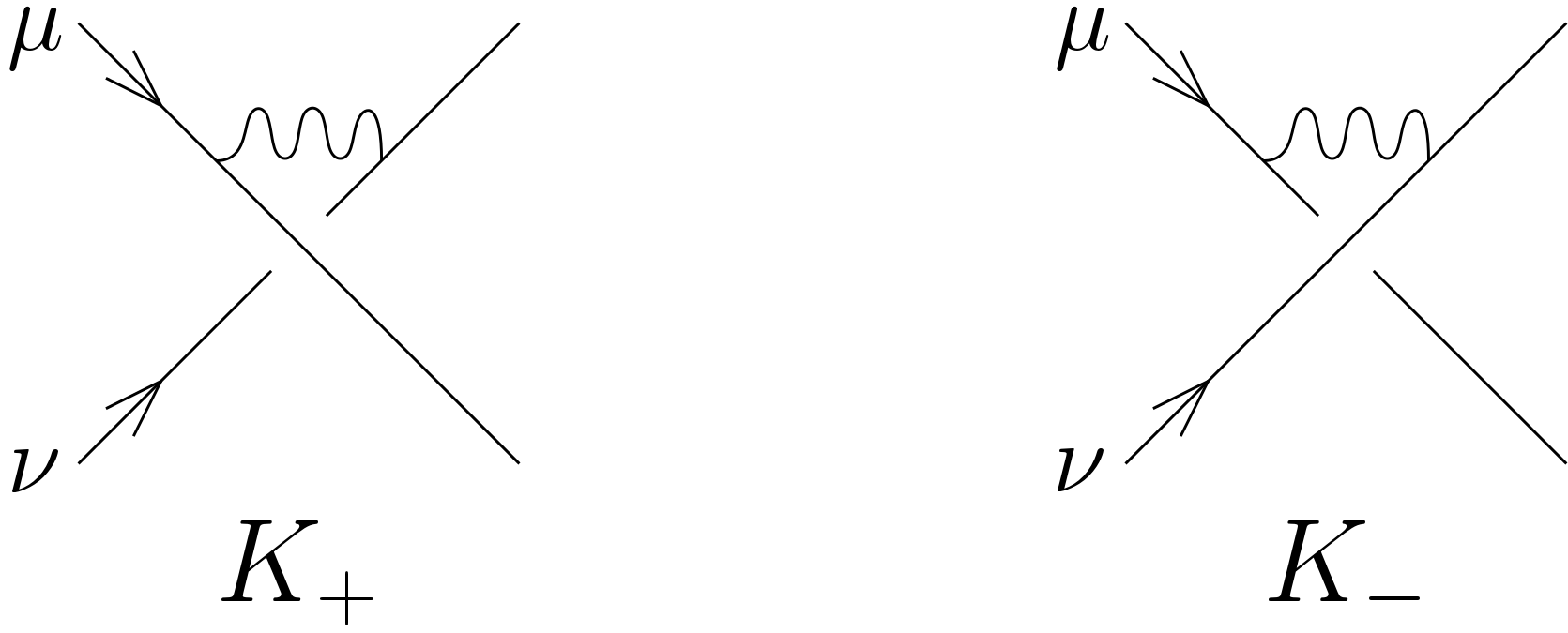}
    \caption{The leading order contributing diagrams for the interaction between Wilson lines corresponding to a gluon exchange at the point of crossing between the lines. There are two types of such crossings: an over-crossing $K_+$ and an under-crossing $K_-$.}
    \label{fig:Cross}
\end{figure}
\noindent Notice that any crossing between two oriented line segments in $\mathbb{R}^2$ can be continuously deformed into one of the two crossing shown in Figure~\ref{fig:Cross}. We will choose as a convention to read a crossing in the following way: the line element coming in from the top left in Figure~\ref{fig:Cross} is associated to the left gauge field in the propagator $P(x,x')=\braket{A(x)A(x')}$ and the line element coming in from the top right is associated to the right gauge field of the propagator. The over-crossing $K_+$ therefore corresponds to the case $x_3>x_3'$ which, considering \eqref{eq:propagator} and \eqref{eq:W1}, means that we should associate to it the $R$-matrix $R_{\mu\nu}=\id+4\pi\hbar r_{\mu\nu}$. Similarly the under-crossing $K_-$ corresponds the case $x_3<x_3'$ so we should associate to it the $\tilde{R}$-matrix $\tilde{R}_{\mu\nu}=\id-4\pi\hbar\tilde{r}_{\mu\nu}$. Recall from Section~\ref{sec:YBE} that the operators $R_{\mu\nu}$ and $\tilde{R}_{\mu\nu}$ swaps the pair of outgoing spin states, which is consistent with the fact that two Wilson lines cross when they interact.

\subsection{Relation to the Braid Group}\label{sec:BG}
It is a well known result (see e.g. \cite{turaev1988yang}) that every $R$-matrix gives rise to a representations of the Artin $n$-strand braid group. Since we have found that interactions in our theory are described by an $R$-matrix, this implies that we can consider Wilson lines to represent braid strands. This will be explained in detail in the present subsection. We start by briefly recalling the concept of braids and the braid group.
\begin{figure}[H]
 	\centering
    \includegraphics[width=0.18\paperwidth]{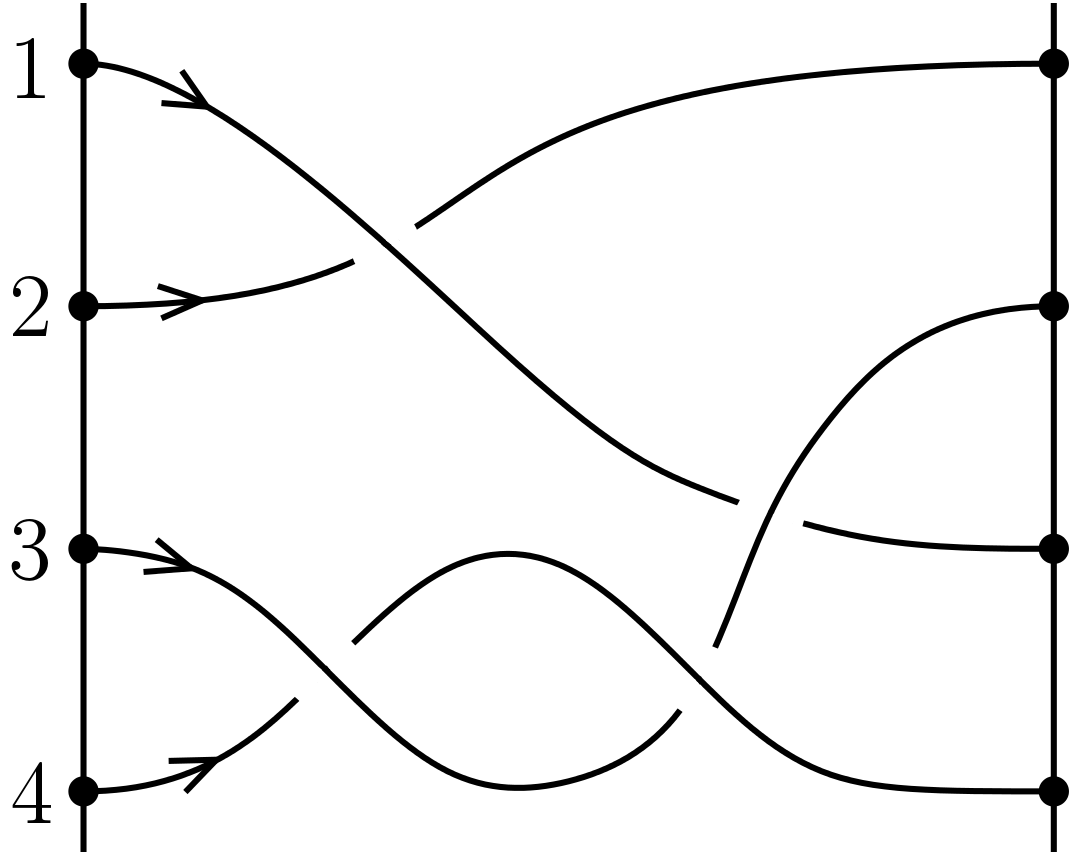}
    \caption{Example of a 4-strand braid.}
    \label{fig:braid}
\end{figure}
\noindent Consider two lines in $\mathbb{R}^3$ parallel to the $y$-axis and with $(x,z)$-coordinates $(0,0)$ and $(1,0)$ respectively, and mark $n$ points on each line. An $n$-strand braid is a set of $n$ non-intersecting curves (strands) connecting the points on the line at $x=0$ with the points on the line at $x=1$ while strictly increasing in the $x$-direction. Similarly to knots, we can represent a braid by its planar projection onto $\mathbb{R}^2$ (corresponding in the above description to the $(x,y)$-plane). A simple example of a planar 4-strand braid is shown in Figure~\ref{fig:braid}.\\

\begin{figure}[H]
 	\centering
    \includegraphics[width=0.5\paperwidth]{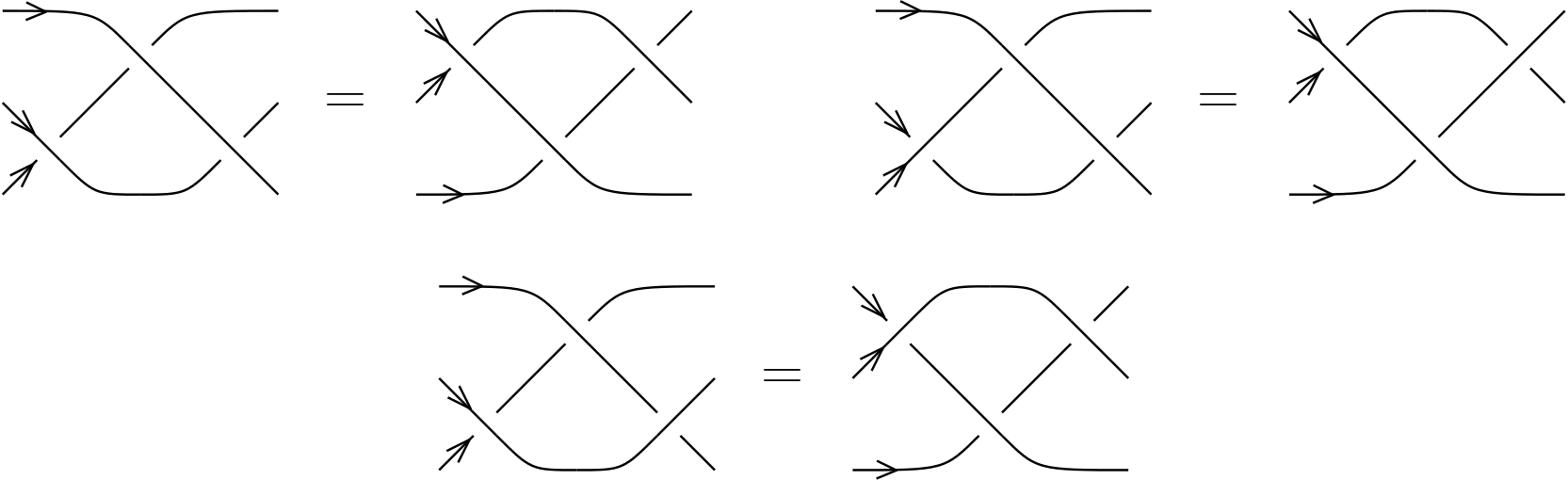}
    \caption{Graphical representation of the defining relations for the Artin braid group: $\sigma_i\sigma_{i+1}\sigma_i=\sigma_{i+1}\sigma_i\sigma_{i+1}$ and its two implications.}
    \label{fig:3dYBE}
\end{figure}

It holds intuitively that any planar braid diagram can be constructed from a series of over-crossings and under-crossing of adjacent strands. This gives rise to the definition of the Artin $n$-strand braid group:
\begin{equation}
\begin{aligned}\label{eq:BG}
B_n=\langle\sigma_1\dots\sigma_{n-1}|\ &\sigma_i\sigma_j=\sigma_j\sigma_i \ \ \text{for} \ |i-j|\geq2,\\ &\sigma_i\sigma_{i+1}\sigma_i=\sigma_{i+1}\sigma_{i}\sigma_{i+1} \ \ \text{for} \ i=1,\dots,n-1\rangle,
\end{aligned}  
\end{equation}
where the graphical interpretation of the generator $\sigma_i$\hspace{2pt}($\sigma_i^{-1}$) is an over\hspace{2pt}(under)-crossing between the braid strands at the $i$'th and $(i+1)$'th position. The first relation in \eqref{eq:BG} is then easily interpreted since the crossing of the strands at $i$ and $i+1$ is obviously independent of the crossing of the strands at $j$ and $j+1$ if $|i-j|\geq 2$. Notice that by multiplying the second relation in \eqref{eq:BG} from the left by $\sigma_i^{-1}$ and from the right by $\sigma_{i+1}^{-1}$ or oppositely we reach the following three equivalent relations:
\begin{equation}
\begin{aligned}\label{eq:braid2}
\sigma_{i}\sigma_{i+1}\sigma_i=\sigma_{i+1}\sigma_i\sigma_{i+1} \ ,& \ \ \sigma_{i}^{-1}\sigma_{i+1}\sigma_i=\sigma_{i+1}\sigma_i\sigma_{i+1}^{-1}\\ \sigma_{i}\sigma_{i+1}\sigma_i^{-1}&=\sigma_{i+1}^{-1}\sigma_i\sigma_{i+1}.
\end{aligned}
\end{equation}
The graphical interpretation of these relations is shown in Figure~\ref{fig:3dYBE}. 
\noindent Starting from a braid, one can obtain a link by closing the strands of the braid, i.e. by connecting the points directly opposite each other. It is a fundamental result in knot theory, known as Alexander's Theorem, that any link can be obtained as the closure of a braid. 
\subsubsection{Wilson Lines and Braids} 
In the definition of a braid given above, we now wish to interpret the braid strands as representing Wilson lines. We thus label each strand by a space of spin states $V_\mu$, $\mu~=~1\dots, n$ transforming under $\mathfrak{sl}_2(\mathbb{C})\oplus\tilde{h}$ and we attach to each crossing between two strands $V_\mu$ and $V_\nu$ an interaction matrix $R_{\mu\nu}$ or $\tilde{R}_{\mu\nu}$ in accordance with the formalism developed in Section~\ref{sec:IntWL}. 
\begin{figure}[H]
 	\centering
    \includegraphics[width=0.35\paperwidth]{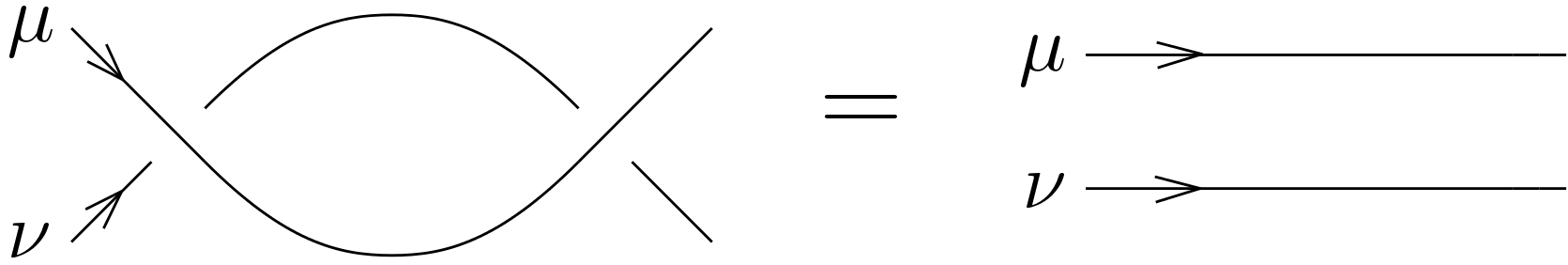}
    \caption{The unitarity relation for Wilson lines.}
    \label{fig:parallel}
\end{figure}
\noindent In order to show that this is a valid interpretation, we start by verifying the unitarity relation illustrated in Figure~\ref{fig:parallel} which states that there is an ``inverse crossing''. In other words, we want to show that the situation where two Wilson lines cross and then cross back without winding around each other is equivalent to the situation where the lines do not cross at all. Written out in equations, we want $\tilde{R}_{\nu\mu}{R}_{\mu\nu}=\id$. We have,
\begin{align}\label{eq:parallel}
\tilde{R}_{\nu\mu}{R}_{\mu\nu}=\left(\id-4\pi\hbar \tilde{r}_{\nu\mu}\right)\left(\id+4\pi\hbar {r}_{\mu\nu}\right)\approx \id-4\pi\hbar \tilde{r}_{\nu\mu}+4\pi\hbar {r}_{\mu\nu}=\id
\end{align}
where the last equality sign follows from the relation $\tilde{r}_{ji}=r_{ij}$. This shows that unitarity is indeed satisfied at leading order in perturbation theory.\\

We now consider the defining relations \eqref{eq:BG} for the Artin braid group. Notice first that $R_{\mu\nu}R_{\lambda\rho}=R_{\lambda\rho}R_{\mu\nu}$ if $\mu,\nu,\lambda,\rho$ are all different, since the $R$-matrices act on different vector spaces. This implies that the first relation in \eqref{eq:BG} is satisfied for Wilson lines and we thus only have left to check the relations in \eqref{eq:braid2}. However, these relations follows immediately from the 3-dimensional generalization of the Yang-Baxter equation \eqref{eq:YBE} which arises with the concept of over-crossings and under-crossings. In fact, using the above unitarity relation, it is relatively straightforward to check that the Yang-Baxter equation extends to the following three equivalent equations which are the analogues of \eqref{eq:braid2}, 
\begin{equation}
    \begin{aligned}\label{eq:3dYBE}
   R_{\mu\nu}R_{\mu\lambda}R_{\nu\lambda}=R_{\nu\lambda}R_{\mu\lambda}R_{\mu\nu}\ ,& \ \ \tilde{R}_{\mu\nu}R_{\mu\lambda}R_{\nu\lambda}=R_{\nu\lambda}R_{\mu\lambda} \tilde{R}_{\mu\nu}, \\ R_{\mu\nu}R_{\mu\lambda}\tilde{R}_{\nu\lambda}&=\tilde{R}_{\nu\lambda}R_{\mu\lambda}R_{\mu\nu},
    \end{aligned}
\end{equation}
The representations of these equations in terms of Wilson line diagrams are exactly the ones given in Figure~\ref{fig:3dYBE}. Thus we have found that interacting Wilson lines in our theory gives rise to a representation of the Artin braid group. We show in Section~\ref{sec:Jones} this actually corresponds to a Hecke algebra representation of the braid group similar to the one used in \cite{jones1990hecke} to construct the Jones polynomials.

\subsection{The Expectation Value of Links}\label{sec:Exp}
Having seen in the previous subsection that open Wilson lines behave like braid strands, we will in the present subsection study the expectation value of links obtained from closing the braid strands. Let $\ket{s_1}$, $\ket{s_2}$ denote the basis vectors of $V$ in the fundamental representation, corresponding to the spin up and spin down states. Then a basis for the total space $V_1\otimes\cdots\otimes V_n$ is given~by
\begin{align}\label{basis}
\{\ket{s_{k_1}}\otimes\cdots\otimes\ket{s_{k_n}} | \ {k_i=1,2} \ , \  i=1,\dots n\}.
\end{align}
Furthermore, denote the total interaction matrix corresponding to a given $n$-strand braid $\alpha$ by $\mathcal{M}_\alpha$, i.e. $\mathcal{M}_\alpha$ is a product of $R$ and $\tilde{R}$ matrices acting on $V_1\otimes\cdots\otimes V_n$. Connecting the braid strands then corresponds to tracing over $\mathcal{M}$ in the basis \eqref{basis} as follows, 
\begin{align}
\braket{W(L)}=\sum_{k_i=1,2}\bra{s_{k_1}}\otimes\dots\otimes\bra{s_{k_n}}\mathcal{M_\alpha}\ket{s_{k_1}}\otimes\dots\otimes\ket{s_{k_n}},
\end{align}
where $L$ is the link obtained from $\alpha$ by closing the braid strands.
\begin{figure}[H]
 	\centering
    \includegraphics[width=0.36\paperwidth]{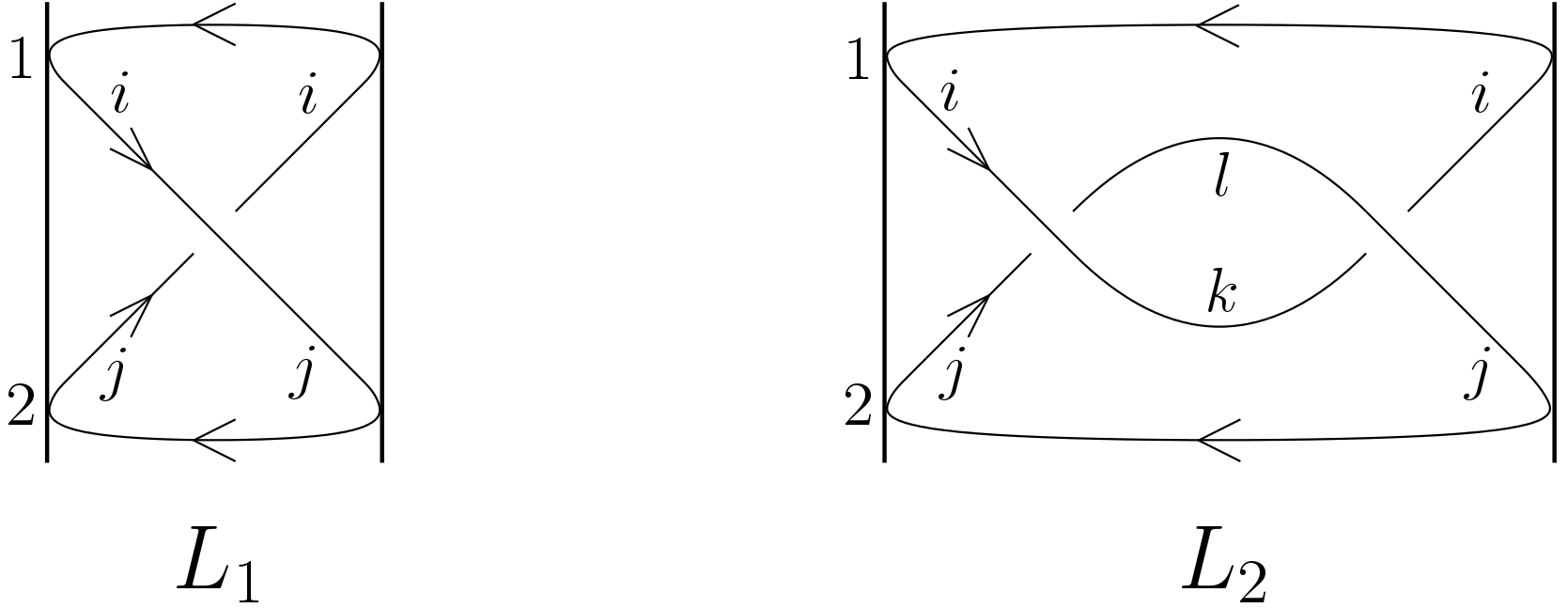}
    \caption{Two examples of Wilson loop diagrams as the closure of braids.}
    \label{fig:Diagrams}
\end{figure}
\noindent In the following, we will determine a general expression for the expectation value of a configuration of Wilson loops at leading order in perturbation theory. As a motivating example, we start by computing the diagrams in Figure~\ref{fig:Diagrams}(a),(b). In the diagrams each line corresponds to a vector space $V_1$ or $V_2$, and each line segment between two crossings is labeled by an $\mathfrak{sl}_2\oplus\tilde{h}$ basis vector. Assigning to each crossing the corresponding $R$-matrix element we obtain,
\begin{equation}\label{eq:exp1}
\begin{aligned}
\braket{W(L_1)}&=\bra{s_{k_1}}\otimes\bra{s_{k_2}}R_{12}\ket{s_{k_1}}\otimes\ket{s_{k_2}}=R^{ij}_{ji}=\delta^i_j\delta^j_i+4\pi\hbar r^{ij}_{ji},
\end{aligned}
\end{equation}
where the indices are summed over. Similarly,
\begin{equation}\label{eq:exp2}
\begin{aligned}
\braket{W(L_2)}&=\bra{s_{k_1}}\otimes\bra{s_{k_2}}R_{12}R_{21}\ket{s_{k_1}}\otimes\ket{s_{k_2}}={R}^{ij}_{kl}{R}^{lk}_{ji}\\&=\left(\delta^i_k\delta^j_l+4\pi\hbar {r}^{ij}_{kl}\right)\left(\delta^l_j\delta^k_i+4\pi\hbar {r}^{lk}_{ji}\right)=\delta^i_i\delta^j_j+8\pi\hbar {r}^{ij}_{ij}.
\end{aligned}
\end{equation}
With the explicit expressions for the generator matrices in \eqref{eq:generators}, \eqref{eq:htilde} we have,
\begin{equation}\label{eq:r-value}
\begin{aligned}
r^{ij}_{ij}&=\tilde{r}^{ij}_{ij}=\frac{1}{4}\Tr(\tilde{h})^2=1  , \\
    r^{ij}_{ji}&=\tilde{r}^{ij}_{ji}=\Tr(ef)+\frac{1}{4}\left(\Tr(h^2)+\Tr(\tilde{h}^2)\right)=2 ,  
\end{aligned}
\end{equation}
and inserting this into \eqref{eq:exp1} and \eqref{eq:exp2} we get, 
\begin{equation}
\begin{aligned}
\braket{W(L_1)}=2+8\pi\hbar, \\ \braket{W(L_2)}=4+8\pi\hbar . 
\end{aligned}
\end{equation}
One can relatively easily convince oneself that, for any $n$-component link of Wilson loops, the term at order zero in $\hbar$ is given by $(\delta^i_i)^n=2^n$. Furthermore, for every crossing between two line segments of the same loop, one must add to the expectation value a term of the form $(2^{n-1})4\pi\hbar r^{ij}_{ji}$ for an over-crossing or $(-2^{n-1})4\pi\hbar \tilde{r}^{ij}_{ji}$ for an under-crossing. Similarly, for every crossing between line segments of distinct loops, one must add a term of the form $(2^{n-2})4\pi\hbar r^{ij}_{ij}$ for an over-crossing or $(-2^{n-2})4\pi\tilde{r}^{ij}_{ij}$ for an under-crossing. Thus, with the values in~\eqref{eq:r-value}, the expectation value of a general configuration of $n$ Wilson loops, $\gamma_1,\dots,\gamma_n$, forming a link $L$ takes the form,
\begin{align}\label{eq:M-corr}
\braket{W(L)}=2^n\Bigg(1+4\pi\hbar\sum_{\alpha=1}^n\big(n^+_{\gamma_\alpha}-n^-_{\gamma_\alpha}\big)+\pi{\hbar}\mathop{\sum_{\alpha,\beta=1}}_{\alpha<\beta}^n\big(n^+_{\gamma_\alpha,\gamma_\beta}-n^-_{\gamma_\alpha,\gamma_\beta}\big)\Bigg), 
\end{align}
where $n^+_{\gamma_\alpha}$($n^-_{\gamma_\alpha}$) is the number of over(under)-crossings in a planar projection of $\gamma_\alpha$ and $n^+_{\gamma_\alpha,\gamma_\beta}$($n^-_{\gamma_\alpha,\gamma_\beta}$) is the number of over(under)-crossings between a line segment of $\gamma_\alpha$ and a line segment of $\gamma_\beta$. We identify in the above expression the writhe number $\omega(\gamma)=(n^+_{\gamma}-n^-_{\gamma})$ and the Gauss linking number $\lk(\gamma_\alpha,\gamma_\beta)=\frac{1}{2}(n^+_{\gamma_\alpha,\gamma_\beta}-n^-_{\gamma_\alpha,\gamma_\beta})$. Defining,
\begin{equation}
\begin{aligned}
\omega(L)\coloneqq\sum_{\alpha=1}^n\omega(\gamma_\alpha) \ , \ \ \ \ \lk(L)\coloneqq\mathop{\sum_{\alpha,\beta=1}}_{\alpha<\beta}^n\lk(\gamma_\alpha,\gamma_\beta),
\end{aligned}
\end{equation}
we can write
\begin{align}\label{eq:exp}
\braket{W(L)}=2^n\Big(1+4\pi\hbar\omega(L)+2\pi{\hbar}\lk(L)\Big)\approx 2^n\exp\left\{4\pi\hbar\left(\omega(L)+\frac{1}{2}\lk(L)\right)\right\}.
\end{align}
The appearance of the linking number in the above equation is in agreement with the well known result (see e.g. \cite{witten1989quantum}) that in abelian Chern-Simons theory the expectation value at first order is given by the Gauss linking number. Notice that the linking number indeed appears from the abelian part of the Lie algebra. The remaining part of \eqref{eq:exp} expresses the framing anomaly of Chern-Simons theory as will be discussed in the following subsection.
\subsection{The Framing Anomaly}\label{sec:FA}
\begin{figure}[H]
    \centering
    \includegraphics[width=0.3\paperwidth]{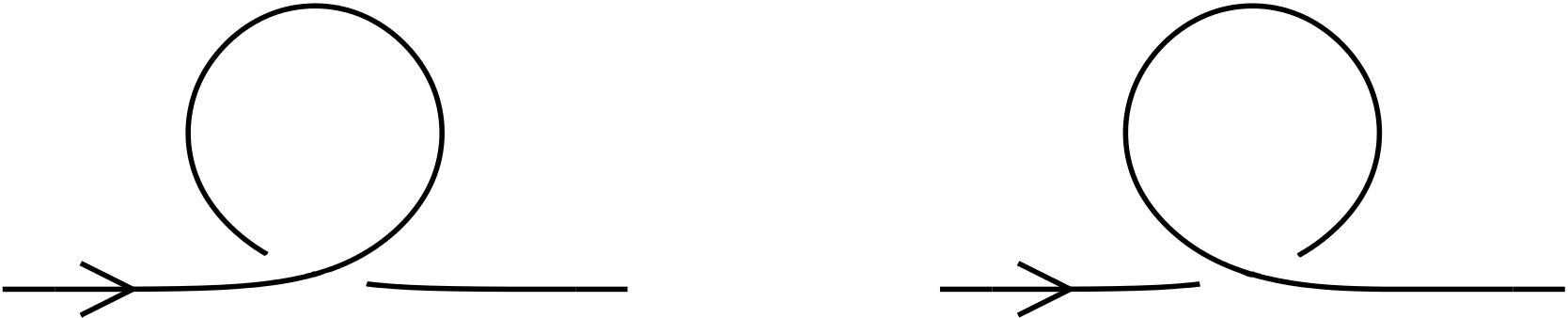}
    \caption{Twisting of a strand with either an under-crossing (left) where $\omega\to\omega-1$, or an over-crossing (right) were $\omega\to\omega+1$.}
    \label{fig:Twist}
\end{figure}

The result in \eqref{eq:exp} implies that the expectation value  of configurations of Wilson loops is not in itself a link invariant. Indeed, a link invariant is defined to be invariant under a set of deformations of the link that one can make without changing its isotopy class. There are three such deformations known as Reidemeister moves. The first Reidemeister move corresponds to twisting a strand of the knot and thereby changing the writhe number by $\omega(L)\to\omega(L)\pm 1$ depending on the type of twist (see Figure~\ref{fig:Twist}). 
\noindent It is evident from the expression \eqref{eq:exp} that under such a twist the expectation value changes according~to
\begin{align}\label{eq:twist}
\braket{W(L)}\to e^{\pm 4\pi\hbar}\braket{W(L)}.
\end{align}  
The remaining two Reidemeister moves are satisfied due to the unitarity relation and the 3-dimensional Yang-Baxter equation which were described in Section~\ref{sec:BG}.\\

This discrepancy in the expectation value under twisting a strand expresses the so called framing anomaly of Chern-Simons theory. In \cite{witten1989quantum} the framing anomaly appears as a consequence of the self-interaction of Wilson loops only being well defined with a choice of framing of the loops and it is found that the expectation value will change under a change of framing. With our method in the present paper we get a well defined expression for the self-interaction without having to introduce a framing. However, as a price, the resulting expression \eqref{eq:exp} is not a link invariant.

\section{Constructing the Jones Polynomials}\label{sec:Jones}
In this section we show in detail how we can recover a specific value of the Jones two-variable polynomials from the expectation value of Wilson loops. The construction that we use is similar to the one originally given by Jones in \cite{jones1990hecke}, namely from Ocneanu's trace acting on a Hecke algebra representation of the Artin braid group. 

\subsection{Hecke Algebra Relation for Wilson Lines}\label{sec:J1}
The Hecke algebra $H_n(q)$ with generators $\{g_i\}_{i=1}^{n-1}$ is obtained from the Artin $n$-strand braid group by adding to \eqref{eq:BG} an additional Hecke algebra relation: 
\begin{align}\label{eq:Hecke}
(g_i-q^{1/2})(g_i+q^{-1/2})=0 \Leftrightarrow g_i=(q^{1/2}-q^{-1/2})+g_i^{-1}  ,
\end{align}
where $q$ is some scalar. The Jones two-variable polynomials were originally constructed from considering a representation of $B_n$ coming from the Hecke algebra\footnote{In the original construction of the Jones polynomials, the generators $\{g_i\}_{i=1}^{n-1}$ was defined to satisfy a slightly modified version of the Hecke algebra relation, given by, $g_i^2=(q-1)g_i+q$. However, the resulting algebra is isomorphic to the one in \eqref{eq:Hecke} obtained by sending $g_i\to q^{-1/2}g_i$. }, and we are therefore interested in checking if a relation similar to the Hecke algebra relation \eqref{eq:Hecke} is satisfied for Wilson lines. \\

\noindent In terms of Wilson line diagrams \eqref{eq:Hecke} corresponds to the following relation: 
\begin{figure}[H]
 	\centering
    \includegraphics[width=0.4\paperwidth]{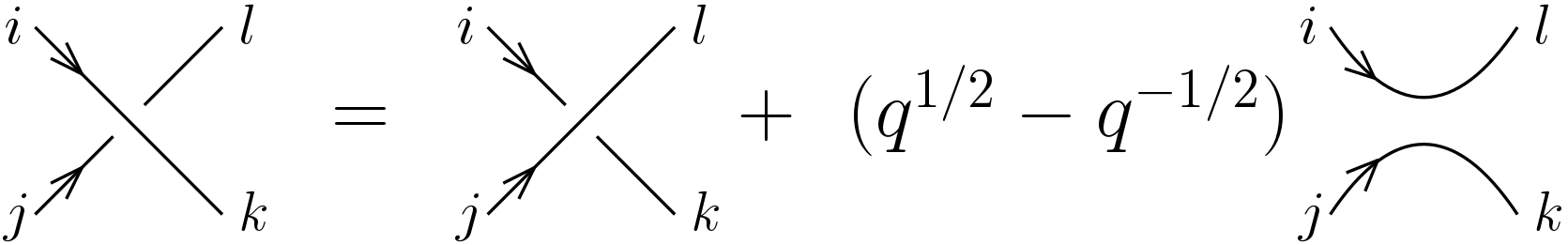}
    \label{Hecke1}
\end{figure}
\noindent where the incoming and outgoing lines are label by spins states. Notice that since the lines in the second term on the right hand side of the equality do not cross, the associated outgoing spin states are swapped relatively to the situation in the other two terms. Thus, the corresponding interaction matrices must differ by a permutation matrix swapping the pair of incoming spin states, i.e. $P^{ij}_{kl}=\delta^i_l\delta^j_k$. As one can easily verify, $P$ can be written in matrix form as  
\begin{align}
P=e\otimes f+f\otimes e+\frac{1}{2}\big(h\otimes h+\tilde{h}\otimes\tilde{h}\big).
\end{align}
By recognizing in the above the solutions of the classical Yang-Baxter equation \eqref{CPsol1} and \eqref{CPsol2}, we find that
\begin{align}\label{eq:Skein}
P=r+\tilde{r}=\frac{1}{4\pi\hbar}(R-\tilde{R}). 
\end{align}
Equivalently, by multiplying both sides of this relation with $P$, we obtain
\begin{align}\label{eq:Skein1}
R_{12}=\tilde{R}_{12}+4\pi\hbar \id.
\end{align}
It is seen that this expression has the form of equation~\eqref{eq:Hecke} when we take $q$ to be $q=e^{4\pi\hbar}$ and expand to first order in $\hbar$. Thus, interacting Wilson lines indeed give rise to a Hecke-algebra representation of the Artin braid group.
  
\subsection{Normalizing the Expectation Value}\label{sec:Ocn}
In \cite{jones1990hecke} the Jones polynomials are obtained from the trace function, $\tr_z$, due to Ocneanu (see \cite{freyd1985new}), acting on $\bigcup_{n=1}^\infty H_n(q)$, which is defined for any $z\in\mathbb{C}$ to satisfy a so called Markov property: $\tr_z(xg_n)=z\tr_z(x)$ for $x\in H_n(q)$. In fact, a Jones polynomial is a polynomial in the variables $q$ and $\lambda$, where $q$ corresponds to the scalar appearing in the definition of the Hecke algebra \eqref{eq:Hecke} and $\lambda$ is a normalisation factor which ensures that $\lambda^{1/2}\tr_z(g_i)=\lambda^{-1/2}\tr_z(g_i^{-1})$. Since the analogue of equation~\eqref{eq:twist} corresponds to the case $\tr_z(g_i^{-1})=z^{-1}$, we take $\lambda=z^{-2}$. Following the definition in \cite{jones1990hecke}, the Jones polynomial for a link $L$ then takes the form,
\begin{align}\label{eq:jones}
V_{L}(q,z)=z^{-e}\tr_z(\pi(\alpha)),
\end{align}
where $\alpha\in B_n$ is any braid whose closure is $L$, $e$ is the exponent sum of $\alpha$ as a word on the $\sigma_i$'s and $\pi$ is the representation of $B_n$ in $H_n(q)$, $\pi(\sigma_i)=g_i$.\\

We notice from the discussion in Section~\ref{sec:FA} that the Markov property of Ocneanu's trace is analogous to the framing anomaly of Wilson loops discussed in~Section~\ref{sec:FA}. Indeed, for a given link $L$, let $x\in H_n(q)$ be the Hecke algebra representation of a braid whose closure is $L$. Then the operation $x\to xg_n$, under which Ocneanu's trace changes by a factor of $z$, corresponds to twisting a strand of $L$ with a twist that adds an over-crossing to $L$. As we known from equation~\eqref{eq:twist}, this increases the total writhe number by 1, which causes the expectation value of $W(L)$ to change by a factor of $e^{\hbar}$. Thus, if we take $q=e^{4\pi\hbar}$ and $\lambda=z^{-2}=e^{-8\pi\hbar}$, the expectation value of Wilson loops can be seen as Ocneanu's trace acting on a Hecke algebra representation of the Artin braid group. In line with the construction in \eqref{eq:jones}, we can therefore obtain a value of the Jones polynomials from the expectation value of Wilson loops by normalising it with a factor that depends on the total writhe number of the link. We define
\begin{align}\label{inv}
X_L\coloneqq e^{-4\pi\hbar \omega(L)}\braket{W(L)}.
\end{align}
It follows from \eqref{eq:twist} that $X_L$ only depends on the isotopy class of $L$ and, according to the above discussion, it corresponds to the specific value of the Jones polynomials given by $V_L(e^{4\pi\hbar},e^{-8\pi\hbar})$. Notice however that, by assumption $V_\bigcirc\equiv1$, where $\bigcirc$ denotes the unknotted circle, and thus $X_L$ differs from $V_L$ by a normalisation.

\subsection{Unknotting Wilson loops}
\begin{figure}[H]
 	\centering
    \includegraphics[width=0.4\paperwidth]{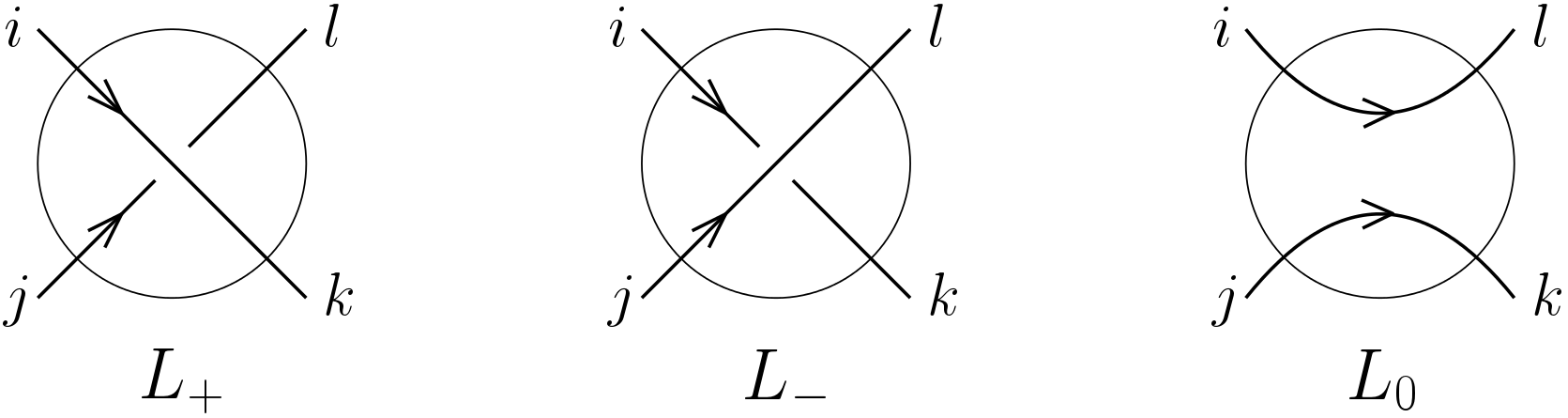}
    \caption{The three skein related links $L_+$, $L_-$ and $L_0$ are identical outside of the encircled area. Incoming and outgoing lines are labeled by spin states of $\mathfrak{sl}_2(\mathbb{C})\oplus\tilde{h}$.}
    \label{fig:Skein}
\end{figure}
A more common way of constructing the Jones polynomial of a given link, is by using that the Jones polynomials are determined uniquely from a so called skein relation. A skein relation in general is a linear relation between the polynomial invariants of links $L_+$, $L_-$ and $L_0$ which differ only at a single crossing where they are as in Figure~\ref{fig:Skein}. Using the Hecke algebra relation \eqref{eq:Hecke} along with the Markov property of Ocneanu's trace one finds that the Jones polynomials satisfy the following skein relation
\begin{align}\label{eq:J-skein}
\lambda^{-1/2}V_{L_+}(q,\lambda)-\lambda^{1/2}V_{L_-}(q,\lambda)=\left(q^{1/2}-q^{-1/2}\right)V_{L_0}(q,\lambda).
\end{align}
Using this relation recursively one can ``unknot'' any link $L$ thus ending up with a set of unlinked unknotted circles which by definition satisfy $V_\bigcirc\equiv 1$. Considering the results in Section \ref{sec:J1} and \ref{sec:Ocn} we expect $X_L$ to be satisfy the skein relation
\begin{align}\label{eq:X-skein}
e^{4\pi\hbar}X_{L_+}-e^{-4\pi\hbar}X_{L_-}=\left(e^{2\pi\hbar}-e^{-2\pi\hbar}\right)X_{L_0}.
\end{align}
Indeed, since the links in Figure~\ref{fig:Skein} are identical outside of the encircled area, the expectation values corresponding to $L_+$, $L_-$ and $L_0$ are related by letting the matrix element corresponding to the crossing inside the encircled area vary between $R^{ij}_{kl}$, $\tilde{R}^{ij}_{kl}$ and $\delta^i_l\delta^j_k$, respectively. We can therefore expand the Hecke algebra relation \eqref{eq:Skein1} to a relation between expectation values as follows 
\begin{align}
\braket{W(L_+)}-\braket{W(L_-)}=\left(e^{2\pi\hbar}-e^{-2\pi\hbar}\right)\braket{W(L_0)}.\label{eq:skein1}
\end{align}
By substituting $\braket{W(L)}=e^{4\pi\hbar\omega(L)}\hspace{1pt}X_L$ into this equation and noting that $\omega(L_+)=\omega(L_0)~+~1$ and $\omega(L_-)=\omega(L_0)-1$, we recover \eqref{eq:X-skein} as expected. This gives a way of recursively unknotting any set of Wilson loops.

\section{Conclusion}
In this paper we have shown that, by imposing boundary conditions in one dimension on the gauge field of a 3-dimensional Chern-Simons theory, the interaction matrix at leading order in perturbation theory takes the form of an $R$-matrix. We argued that this result allows us to recover the Jones two-variable polynomials for specific values of the variables from the expectation value of Wilson loops using a construction analogous to the one originally given by Jones. Our results therefore give new insight into the relation between Chern-Simons theory and knot theory. We have been working only to leading order in perturbation theory and so it would be interesting, as a further investigation, to verify that the constructions can be generalized to higher orders.

\section*{Acknowledgements}
I would like to thank my supervisor Kevin Costello for helpful guidance. I also wish to thank Victor~Py for useful comments along the way. This research was supported in part by Perimeter Institute for Theoretical Physics. Research at Perimeter Institute is supported by the Government of Canada through the Department of Innovation, Science, and Economic Development, and by the Province of Ontario through the Ministry of Research and Innovation.

\end{document}